\documentclass[12pt]{article}
\usepackage{pic04,epsfig}

\begin{document}

\title{\bf EXPERIMENTAL RESULTS AND THEORETICAL DEVELOPMENTS OF MUON g-2}
\author{
H.~Deng$^{11}$,
G.W.~Bennett$^{2}$,
B.~Bousquet$^{9}$,
H.N.~Brown$^2$,
G.~Bunce$^2$,\\
R.M.~Carey$^1$,
P.~Cushman$^{9}$,
G.T.~Danby$^2$,
P.T.~Debevec$^7$,
M.~Deile$^{11}$,\\
S.K.~Dhawan$^{11}$,
V.P.~Druzhinin$^3$,
L.~Duong$^{9}$,
F.J.M.~Farley$^{11}$,
G.V.~Fedotovich$^3$,\\
F.E.~Gray$^7$,
D.~Grigoriev$^3$,
M.~Grosse-Perdekamp$^{11}$,
A.~Grossmann$^6$,
M.F.~Hare$^1$,\\
D.W.~Hertzog$^7$,
X.~Huang$^1$,
V.W.~Hughes$^{11}$,
M.~Iwasaki$^{10}$,
K.~Jungmann$^5$,\\
D.~Kawall$^{11}$,
B.I.~Khazin$^3$,
F.~Krienen$^1$,
I.~Kronkvist$^{9}$,
A.~Lam$^1$,\\
R.~Larsen$^2$,
Y.Y.~Lee$^2$,
I.~Logashenko$^{1,3}$,
R.~McNabb$^{9}$,
W.~Meng$^2$,\\
J.P.~Miller$^1$,
W.M.~Morse$^2$,
D.~Nikas$^2$,
C.J.G.~Onderwater$^{7}$,
Y.~Orlov$^4$,\\
C.S.~\"{O}zben$^{2,7}$,
J.M.~Paley$^1$,
Q.~Peng$^1$,
C.C.~Polly$^7$,
J.~Pretz$^{11}$,\\
R.~Prigl$^{2}$,
G.~zu~Putlitz$^6$,
T.~Qian$^{9}$,
S.I.~Redin$^{3,11}$,
O.~Rind$^1$,\\
B.L.~Roberts$^1$,
N.~Ryskulov$^3$,
Y.K.~Semertzidis$^2$,
P.~Shagin$^9$,
Yu.M.~Shatunov$^3$,\\
E.P.~Sichtermann$^{11}$,
E.~Solodov$^3$,
M.~Sossong$^7$,
L.R.~Sulak$^{1}$,
A.~Trofimov$^1$,\\
P.~von~Walter$^6$,
and
A.~Yamamoto$^8$.
\\
(Muon $(g-2)$ Collaboration)
}

\maketitle

\vspace{1cm}
\begin{center}
{\em
$\,^1$Department of Physics, Boston University, Boston, Massachusetts 02215\\
$\,^2$Brookhaven National Laboratory, Upton, New York 11973\\ 
$\,^3$Budker Institute of Nuclear Physics, Novosibirsk, Russia\\
$\,^4$Newman Laboratory, Cornell University, Ithaca, New York 14853\\
$\,^5$ Kernfysisch Versneller Instituut, Rijksuniversiteit
Groningen, NL 9747\,AA Groningen, The Netherlands\\
$\,^6$ Physikalisches Institut der Universit\"at Heidelberg, 69120 
Heidelberg, Germany\\
$\,^7$ Department of Physics, University of Illinois at Urbana-Champaign,
Illinois 61801\\
$\,^8$ KEK, High Energy Accelerator Research Organization, Tsukuba,
Ibaraki 305-0801, Japan\\
$\,^{9}$Department of Physics, University of Minnesota,Minneapolis,
Minnesota 55455\\
$\,^{10}$ Tokyo Institute of Technology, Tokyo, Japan\\
$\,^{11}$Department of Physics, Yale University, New Haven, Connecticut 06520
}
\end{center}
\newpage

\baselineskip=14.5pt
\begin{abstract}
The anomalous magnetic moments of both positive and negative muon have been
meausred to a precision of 0.7 parts per million (ppm) at the Brookhaven
Alternating Gradient Synchrotron. The results are over an order of magnitude
more precise than the previous measurement of the muon. The results 
$a_{\mu^+}=11~659~204(7)(5) \times 10^{-10}~(0.7\,{\rm ppm})$ and 
$a_{\mu^-}=11~659~214(8)(3) \times 10^{-10}~(0.7\,{\rm ppm})$, where the first
uncertainty is statistical and the second is systematic, are consistent with
each other. The average for the muon anomaly is 
$a_{\mu}({\rm exp})=11~659~208(6) \times 10^{-10}~(0.5\,{\rm ppm})$.   
The standard model value of $a_{\mu}$ is calculated to a precision of $0.6$ppm,
which is dominated by uncertainty of hadronic contributions. There is a
significant disagreement between the hadronic contribution calculated from 
the hadronic cross section of $e^+e^-$ collision and that from hadronic
$\tau$ decay. The difference between experimental and theoretical values
of $a_\mu$ is $2.4 \sigma$ for $e^+e^-$ data or $0.9 \sigma$ for $\tau$ decay
data.
\end{abstract}
\newpage

\baselineskip=17pt

\section{\bf Introduction}
The gyromagnetic factor of a particle is defined as
\begin{equation}
g=\left(\frac{\mu}{e\hbar/2mc}\right)/\left(\frac{S}{\hbar}\right),
\label{eq:defg}
\end{equation}
where $\mu$, $S$, $m$, and $e$ are the magnetic moment, the spin,
the mass, and the charge of that particle.
For a pointlike lepton Dirac theory gives $g=2$. Virtual radiative
corretions cause the $g$ value for leptons to differ from 2. The anomalous
$g$-value is defined as $a=(g-2)/2$.

The anomalous $g$-value of the electron and muon, $a_e$ and $a_{\mu}$,
played an important role in
the development of particle physics and continue to serve as fundamental
quantities for testing the validity of the Standard Model and putting
stringent constraints on speculative theories beyond the Standard Model.
The value of $a_e$ is measured to the precision of 4 parts per billion (ppb).
This value is currently the most precise measurement
of fine structure constant $\alpha$.

For the muon, because the muon mass $m_{\mu}$ is larger than the electron
mass $m_e$, the
contribution to $a_{\mu}$ from heavier particles such as hadrons or weak
vector bosons through virtual processes is greater than that to $a_e$
typically by a factor of
$(m_{\mu}/m_e)^2 \simeq 4 \times 10^4$. This sensitivity enhancement of
$a_{\mu}$ applies generally also for postulated new particles such as
supersymmetric particles. Hence, the measurement of $a_\mu$ provides us more
information than that of $a_e$ though $a_\mu$ is less well known than $a_e$.

The muon anomalous magnetic moment was measured three times in
CERN \cite{cern1,cern2,cern3}, and the precision of the last CERN measurement
is $7.2$ ppm.
The muon $g-2$ experiment at Brookhaven National Laboratory (E821) measures
the $a_\mu$ of both the positive and the negative muon to $0.7$ ppm. Two values
agree with each other and the precision of the combined value of $a_\mu$
is $0.5$ ppm\cite{1997,1998,1999,2000,2001}.

\section{\bf Principle and setup of the experiment}

For a muon moving in a uniform magnetic field $\vec{B}$, which is
perpendicular to the muon spin direction and to the plane of the orbit, and
with an electric field $\vec{E}$, both the spin and the momentum precess.
The angular frequency difference,
$\omega_a$, between the spin precession frequency
$\omega_s$ and the cyclotron frequency $\omega_c$ is given by
\begin{equation}
\vec{\omega}_a=\vec{\omega}_s-\vec{\omega}_c=-\frac{e}{m_{\mu}c}
\left[a_{\mu}\vec{B}-\left(a_{\mu}-\frac{1}{\gamma^2-1}\right)\vec{\beta}
\times \vec{E}\right]
\label{eq:nomagic}
\end{equation}
for $\vec{\beta}\cdot \vec{B}=0$ and $\vec{\beta} \cdot \vec{E}=0$,
where $m_\mu$ and $e$ are the mass and charge of the muon, $c$ is the speed of
light, $\beta$ is the ratio of speed of the muon and the speed of light, and
$\gamma=1/\sqrt{1-\beta^2}$. In muon $g-2$ experiment,
the dependence of $\omega_a$ on the electric field is eliminated by storing
muons with the ``magic'' $\gamma=29.3$ such that
\begin{equation}
a_{\mu}-\frac{1}{\gamma^2-1}=0.
\label{eq:magic}
\end{equation}
This corresponds to a muon momentum $p=3.09$~GeV/c. Hence, the
measurement of $\omega_a$ and $B$ determines $a_{\mu}$,
\begin{equation}
a_{\mu}=\frac{\omega_a}{\frac{e}{m_{\mu}c}B}.
\label{eq:amdef0}
\end{equation}
In our experiment, the magnetic field is measured by an NMR system which
is calibrated with respect to the free proton NMR angular frequency
$\omega_p$ as
\begin{equation}
B=\frac{\hbar \omega_p}{2\mu_p},
\label{eq:Bfreep}
\end{equation}
where $\mu_p$ is the magnetic moment of the free proton.
By combination with \ref{eq:amdef0}, \ref{eq:Bfreep} and
using muon spin $S/\hbar=1/2$ and the definition of $g_\mu=2(1+a_\mu)$,
\begin{equation}
a_{\mu}=\frac{\omega_a}{\frac{e}{m_\mu c}\frac{\hbar \omega_p}{2\mu_p}}
=\frac{\omega_a}{\frac{4\mu_\mu}{\hbar g_\mu}\frac{\hbar \omega_p}{2\mu_p}}
=\frac{\omega_a/\omega_p}{\frac{\mu_\mu}{(1+a_\mu)\mu_p}}
=\frac{\omega_a/\omega_p}{\mu_\mu/\mu_p}(1+a_\mu),
\end{equation}
so that
\begin{equation}
a_\mu=\frac{\omega_a/\omega_p}{\mu_\mu/\mu_p - \omega_a/\omega_p}.
\label{eq:amdef}
\end{equation}
The ratio $\mu_\mu/\mu_p = 3.183~345~39(10)$\cite{groom}, as determined
from muonium spectrocopy \cite{muonium}.

The frequency $\omega_a$ is measured by counting high energy decay positrons.
In the muon rest frame, parity violation in the decay
$\mu^+ \rightarrow e^+ \nu_e \bar{\nu}_{\mu}$ causes $e^+$ to be emitted
preferentially along the muon spin direction. Positrons emitted
along the muon momentum direction get the largest Lorentz boost and have
highest energy in the laboratory frame.
Hence, the angle between the muon momentum and muon spin determines
the number of high energy decay positrons. While the muon spin precesses with
angular frequency $\omega_s$ and the muon momentum precesses with angular
$\omega_c$ in the magnetic field, the number of decay positrons with
energy greater than some threshold energy $E$ oscillates with the angular
frequency $\omega_a=\omega_s-\omega_c$ and is ideally described by
\begin{equation}
N(t)=N_0(E)e^{-t/(\gamma \tau)}\{1+A(E)\sin[\omega_at+\phi_a(E)]\}
\end{equation}
in which $\gamma \tau$ is the dilated muon life time, $A$ is the asymmetry,
$\phi$ is the phase of the oscillation. Both $A$ and $\phi$
depend on the energy threshold of the decay positrons. However, the
frequency $\omega_a$ does not depend on $E$.

\begin{figure}[ht]
\center
\epsfig{file=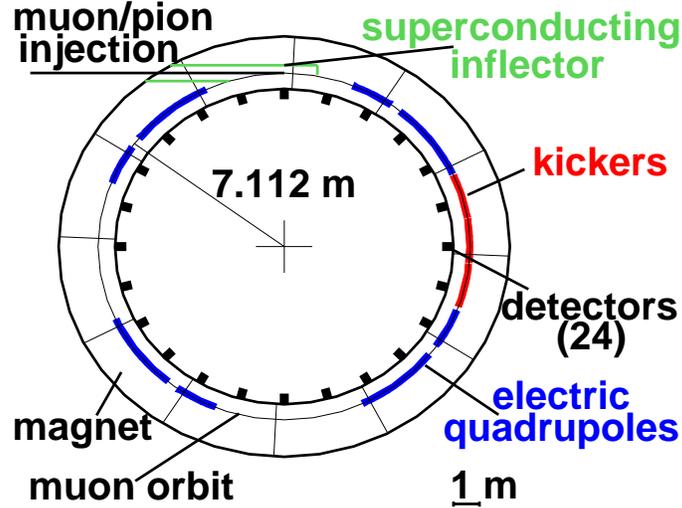, width=0.6\textwidth}
\caption{\it General arrangement of the muon $g-2$ experiment.}
\label{fig:ring}
\end{figure}

The general arrangement of the muon $g-2$ experiment is shown in
Fig.~\ref{fig:ring}.
The proton beam from the Alternating Gradient Synchrotron(AGS)
strikes a nickel target to produce pions. The pions are captured into a
secondary beamline where a fraction of about $50\%$ decays to muons.
We select the high energy muons which come from the
forward decay. They are polarized due to the parity violation in the decay
$\pi^+ \to \mu^+ \nu_\mu$. The muons are injected through a superconducting
inflector\cite{inflector} 
into the muon storage ring with 1.45~T uniform magnetic field\cite{danby}
which is measured by a NMR system\cite{prigl} relative to the free proton
NMR frequency. A pulsed magnetic kicker\cite{kicker}
at about $90^\circ$ from the injection point gives
muons a 10~mrad deflection which places the muons onto stored orbits.
The electrostatic quadrupole\cite{quadrupole}
fields provide vertical focusing. The decay
positrons are detected by 24 lead/scintillating fiber electromagnetic
calorimenters\cite{calo} read out by waveform digitizers. Both the waveform digitizer
and NMR clocks were phase locked to the Loran C frequency signal.

\section{\bf Data Analysis}

The data analysis in the experiment is divided into two independent parts,
$\omega_p$ and  $\omega_a$ analyses. The absolute values of $\omega_p$ and
$\omega_a$ from analyses were modified by arbitrary offsets. No one knew
both absolute values of $\omega_p$ and $\omega_a$.
The values of $R=\omega_a/\omega_p$ and
then $a_\mu$ were evalueated only after both analyses had been finalized.

\subsection{$\omega_p$ analysis}

The measurements of the magnetic field at sub-ppm lever is done with
NMR probes and is expressed relative to the free proton NMR frequency
$\omega_p$.

The NMR frequencies are measured with respect to a standard probe, which is a
spherical water sample. The calibration of the standard probe with
respect to the free proton is obtained from other experiments\cite{fei}.

A trolley with 17 NMR probes was used to measure the field around the
storage ring. The 17 trolley probes are calibrated with
respect to the standard probe through a secondary probe called plunging probe,
which can be mounted on a stand and moved in radial and vertical direction
in vacuum chamber at one azimuthal location. The calibrations were done by
comparing the NMR readings of the plunging probe and the trolley probes taken
at the same location. The errors come from the position uncertainties of the
trollye probes and of the plunging probe, and from the B field inhomogeneity,
Large magnetic sextupole moments and field gradient in azimuth direction
were introduced to measure the relative position of active volume of
trolley probes and the plunging probe. The effect on the trolley probe
calibration from temperature, power supply and magnetic field
were also carefully studied.

\begin{figure}[ht]
\center
\begin{minipage}{0.45\textwidth}
\epsfig{file=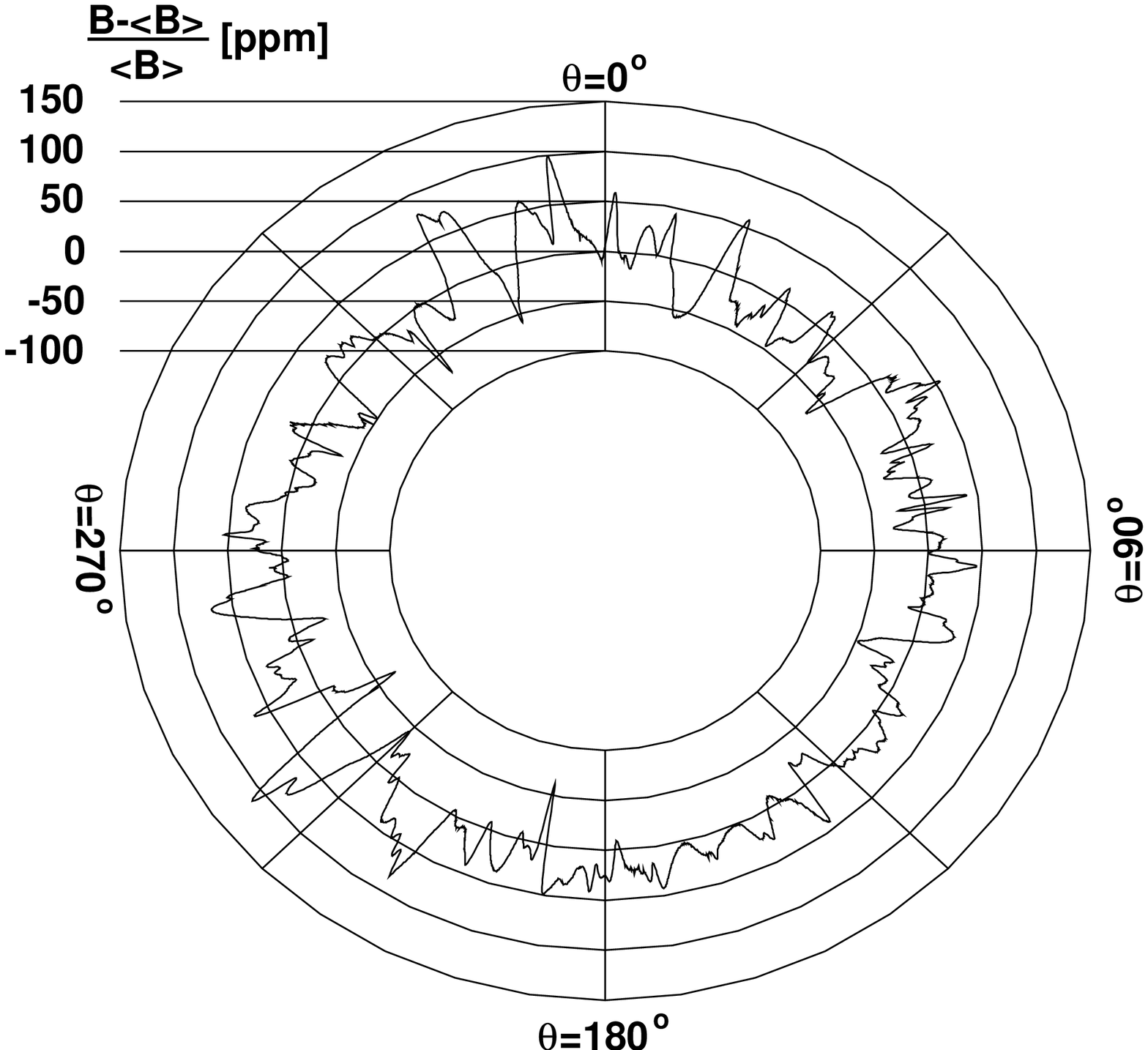, height=5cm}
\end{minipage}
\begin{minipage}{0.51\textwidth}
\epsfig{file=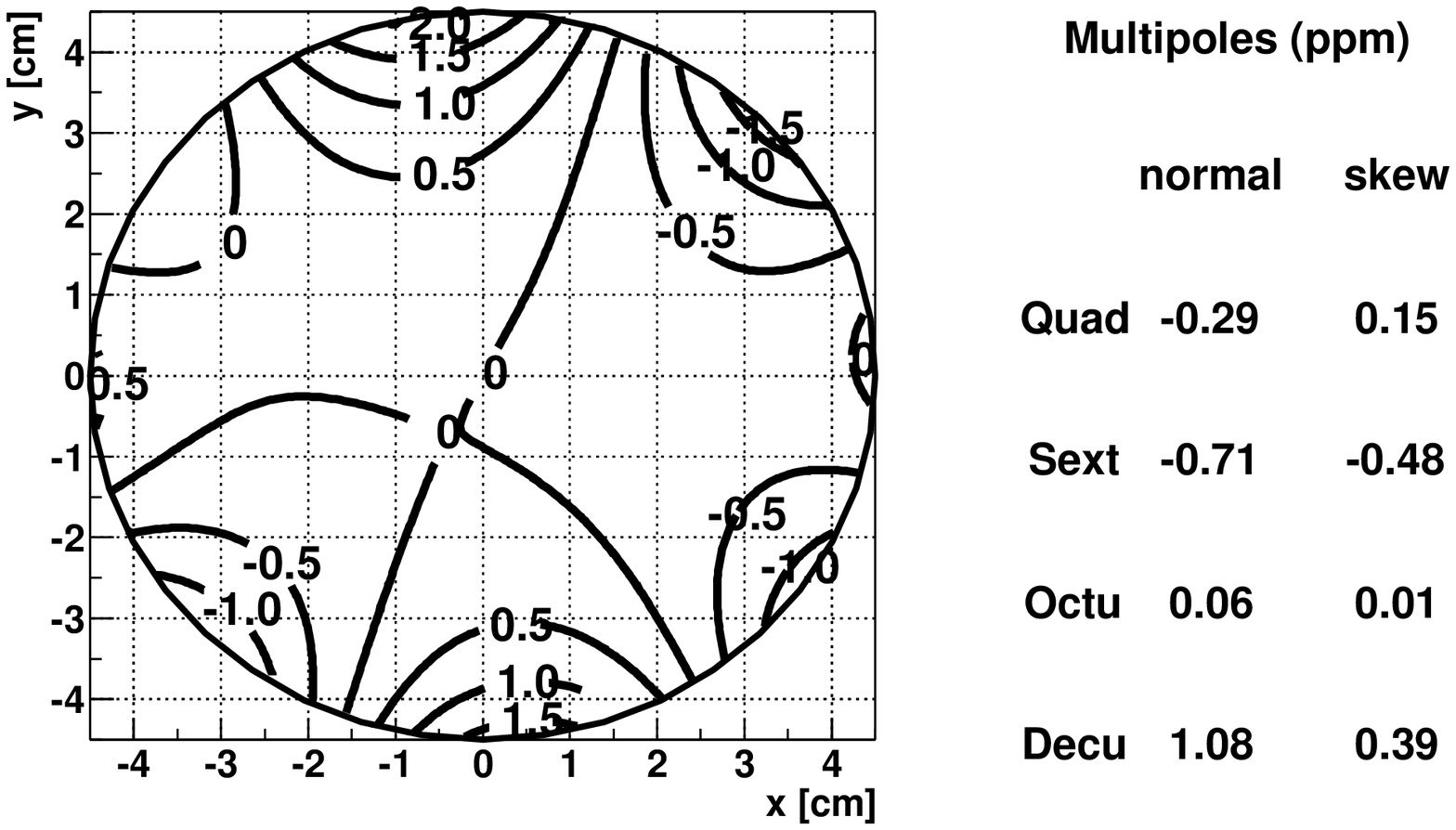, height=5cm}
\end{minipage}
\caption{\it The left plot shows the magnetic field measured with the trolley
center probe. And the right plot shows the half ppm contour plot and the 
multipoles of the azimuthal
average of the field measured in a trolley run.}
\label{fig:multipole}
\end{figure}

We measured the magnetic field with trolley about twice per week. One of
the measurements of the magnetic field with the trolley center probe
is shown vs. azimuth in Figure \ref{fig:multipole}.
For most of the ring, the magnetic field varies about $\pm 50$~ppm.
The field measurements are averaged over azimuth for all 17
probes individually. Since the field component in radial and azimuthal
direction are much smaller than that in vertical direction, the field
can be expanded in 2-dimension
\begin{equation}
B(x,y)=B(r,\theta)=B_0+\sum_{i=1}^{4}a_i\left(\frac{r}{r_0}\right)^i
\cos(i\theta)+\sum_{j=1}^{4}b_i\left(\frac{r}{r_0}\right)^j\sin(j\theta),
\label{eq:multi_expansion}
\end{equation}
where $x$ is the radial direction in our ring, and $(r, \theta)$ are polar
coordinates. The center of the storage region is at $r=0$ and the direction
pointing outward is $\theta=0$. The value of $r_0$ is set to be 4.5~cm.
A half ppm contour plot and the multipoles of the azimuthal average of the
field measured in one trolley run are shown in Figure~\ref{fig:multipole}.

The field between trolley runs was measured by a fixed probe system of 378
NMR probes. More than 120 probes are used in the analysis because of their
good qualities. The average measurements from those fixed probes 
during the trolley run is calibrated with respect to the trolley measurements.

The muon distribution is taken into account by using the field value at the
center of the muon distribution. The radial muon distribution is measured by
fast rotation with uncertainty of 1~mm. The vertical muon distribution is
symetric around the central plane determined by electric quadruples, and
the uncertainty is 2~mm.

The contributions to the total systematic error from different sources are
given in Table~\ref{tb:Berror} for 2001 run. 
And the result for
$\omega_p$ weighted by the muon distribution in 2001 is found to be
\begin{equation}
\omega_p/(2\pi)=61~791~595(15)~{\rm Hz}~(0.17{\rm ppm})
\end{equation}

\begin{table}[ht]
\caption {Systematic uncertainties for $\omega_p$ analysis of the 2001 data}
\vspace{0.3cm}
\center
\begin{tabular}{||l|l||} \hline
Source of errors & Size [ppm] \\
\hline
Absolute calibration of standard probe & $0.05$\\
Calibration of trolley probes &  $0.09$ \\
Trolley measurements of $B_0$  & $0.05$ \\
Interpolation with fixed probes & $0.07$ \\
Uncertainty from muon distribution & $0.03$\\
Others $^\dagger$ & $0.10$\\
\hline
Total Systematic Error on $\omega_p$ & $0.17$ \\
\hline
\end{tabular}

$^\dagger$ Higher multipoles, trolley temperature and voltage
response, and eddy currents from the kickers, and time-varying stray fields..
\label{tb:Berror}
\end{table}

\subsection{$\omega_a$ analysis}

\begin{figure}[ht]
\center
\epsfig{file=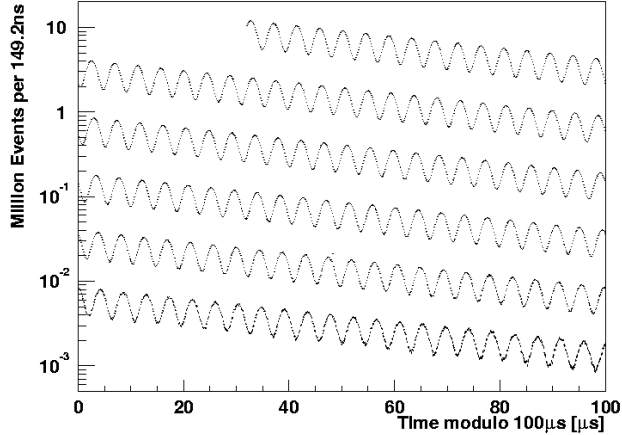, width=0.6\textwidth}
\caption{\it The time spectrum of electrons}
\label{fig:wiggle}
\end{figure}

In 2001 run, We collected about 4 billions of electrons $32\mu s$ after
the injection and with reconstructed energy larger than 1.8GeV. The detection
time of electrons are randomized by cyclotron period to eliminate the bunch
structure at early time. And the time spectrum after randomization is shown
in Fig.~\ref{fig:wiggle}.

The leading characteristics of the time spectrum are those of muon decay and
the muon g-2 oscillation, $\omega_a$,  as shown in Eq.~\ref{eq:5par}
\begin{equation}
N(t)=N_0 e^{-t/\gamma \tau}[1+Acos(\omega_a t+ \phi)]
\label{eq:5par}
\end{equation}
Additional effects includes pulse overlap, detector gain stability,
muon loss and coherent betatron oscillation (CBO).
Among those, CBO in horizontal plane
has the largest effect on the determination of $\omega_a$. CBO is
caused by injecting muons through narrow aperture of the inflector into
larger aperture of the strorage ring. Since the calorimeter acceptance
varies with the decay position of muons, the time and energy
spectra of decay electrons are modulated with the CBO frequency.
The beating frequency between CBO and $\omega_a$ is very close to
$\omega_a$. Hence, the CBO has large effect on determination of $\omega_a$.
Since the phase of CBO oscillation varies from 0 to $2\pi$ around
the ring, its effect in the sum of spectra from all detectors is significantly
smaller than that in any individual detector. The CBO effect is also
accounted in the fitting functions.
In 2001, we ran our experiment at two values of CBO frequency, one with CBO
frequency larger than $2 \times \omega_a$ and the other with CBO frequecy
smaller than $2 \times \omega_a$. Fitting the sum spectrum with function 
taking into account the CBO gave small systematic error for both cases. 

Two approaches were used to obtain the value of $\omega_a$. In one approach,
we fit the time spectrum directly with some multi-parameter function. The function
includes Eq.~\ref{eq:5par} and the effects from gain change, CBO, muon losses.
The fitting function is slightly different for different analyzer. The other
approach is called ratio method. The data are randomly assigned to four
statistically independent subsets $n_1$ to $n_4$. The subsets are rejoined
in $u(t)=n_1(t)+n_2(t)$ and $v(t)=n_3(t-\tau/2)+n_4(t+\tau/2)$, where $\tau$
is an estimate of the $(g-2)$ period, and then combined to form the time
spectrum $r(t)=[u(t)-v(t)]/[u(t)+v(t)]$. The $(g-2)$ rate modulation of
$v$ is $180^\circ$ degrees out of phase compared to that of $u$, and to
sufficient precision $r(t)$ can be described by
\begin{equation}
r(t)=A\sin(\omega_at+\phi_a)+(\tau_a/16\tau)^2.
\end{equation}
where $\tau_a$ is the $(g-2)$ period and $\tau$ is the dilated lifetime of
muons.
The ratio $r(t)$ is largely insensitive to change
of observed counts on time scales larger than $\tau_a$. However, CBO effect
still has to be taken into account in ratio method.

We completed four independent analyses. They all agree with one another within
the expected variation. The combined result
\begin{equation}
\omega_a=229~073.59(15)(5){\rm Hz~(0.7ppm)}, 
\end{equation}
where the first uncertainty is
statistical uncertainty and the second is the combined systematic uncertainty. 
And the systematic errors are listed in Table~\ref{tb:omegaa}.

\begin{table}[ht]
\caption {Systematic uncertainties for $\omega_a$ analysis of the 2001 data}
\vspace{0.3cm}
\center
\begin{tabular}{||l|l||} \hline
Source of errors & Size [ppm] \\
\hline
Coherent betatron oscillation & $0.07$\\
Pileup &  $0.08$ \\
Gain change  & $0.12$ \\
Lost muons & $0.09$ \\
Others $^\dagger$ & $0.11$\\
\hline
Total Systematic Error on $\omega_a$ & $0.21$ \\
\hline
\end{tabular}

$^\dagger$ AGS background, timing shifts, E field and vertical oscillations,
beam debunching/randomization, binning and fitting procedure.
\label{tb:omegaa}
\end{table}

\subsection{results}

The value of $a_\mu$ was determined after the analyses of $\omega_p$ and
$\omega_a$ had been finalized,
\begin{equation}
a_{\mu^-}=11~659~214(8)(3) \times 10^{-10} (0.7~{\rm ppm})
\end{equation}
in which the correlations between the systematic uncertainties have been
taken into account.

\begin{figure}[ht]
\center
\epsfig{file=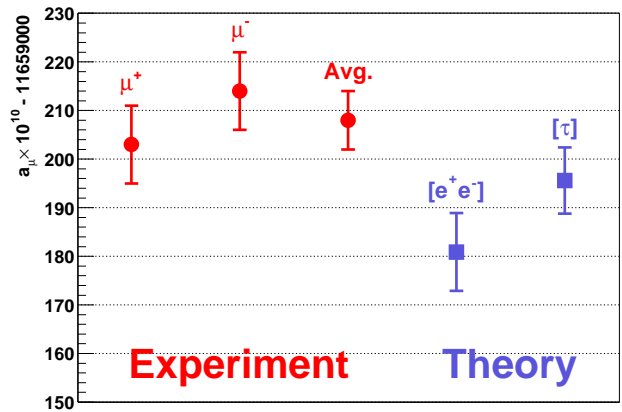, width=0.6\textwidth}
\caption{\it The experimental results and the theoretical prediction of $a_\mu$.}
\label{fig:finalresult}
\end{figure}

In Eq.~\ref{eq:amdef}, $\mu_\mu / \mu_p$ is measured by \cite{muonium} to 
high precision for $\mu^+$ only. Hence $a_{\mu^+}$ and $a_{\mu^-}$
cannot be done directly. Instead, we compared $R=\omega_a/\omega_p$ values
from $\mu^+$ ad $\mu^-$. The difference is
\begin{equation}
\epsilon=\frac{1}{2}\frac{R_{\mu^+}-R_{\mu^-}}{R_{\mu^+}+R_{\mu^-}}=-9.4\pm9.7 \times 10^{-7}
\end{equation}
and is consistent with 0. Assuming CPT, we combine the results from $\mu^+$
and $\mu^-$ and obtain
\begin{equation}
a_\mu=11~659~208(6) \times 10^{-10}
\end{equation}
Fig.~\ref{fig:finalresult} shows the measurements
and the theoretical prediction of $a_\mu$.

\section{\bf Theory}
\label{theory}
In the Standard Model, the contributions to $a_\mu$ are divided into three
types, QED, hadronic and weak. 
\begin{equation}
a_\mu({\rm SM})=a_\mu({\rm QED})+a_\mu({\rm had})+a_\mu({\rm weak}).
\end{equation}

The QED contribution has the largest contribution but least uncertainty. The
calculation up to $5^{th}$ order of $\alpha/(2\pi)$ gives
\begin{equation}
a_\mu({\rm QED})=11~658~470.56(29) \times 10^{-10} \cite{kinoshita}
\end{equation}
A recent recalculation of the terms of $4^{th}$ order of $\alpha/(2\pi)$
increases the QED contribution to
\begin{equation}
a_\mu({\rm QED})=11~658~471.94(14) \times 10^{-10} \cite{kinoshitarecent}
\end{equation}
The weak contribution has been calculated to the second order and is evaluated
as
\begin{equation}
a_\mu({\rm weak})=15.4(2)\times 10^{10} \cite{czarnecki}
\end{equation}

\begin{figure}[ht]
\center
\begin{minipage}{0.48\textwidth}
\epsfig{file=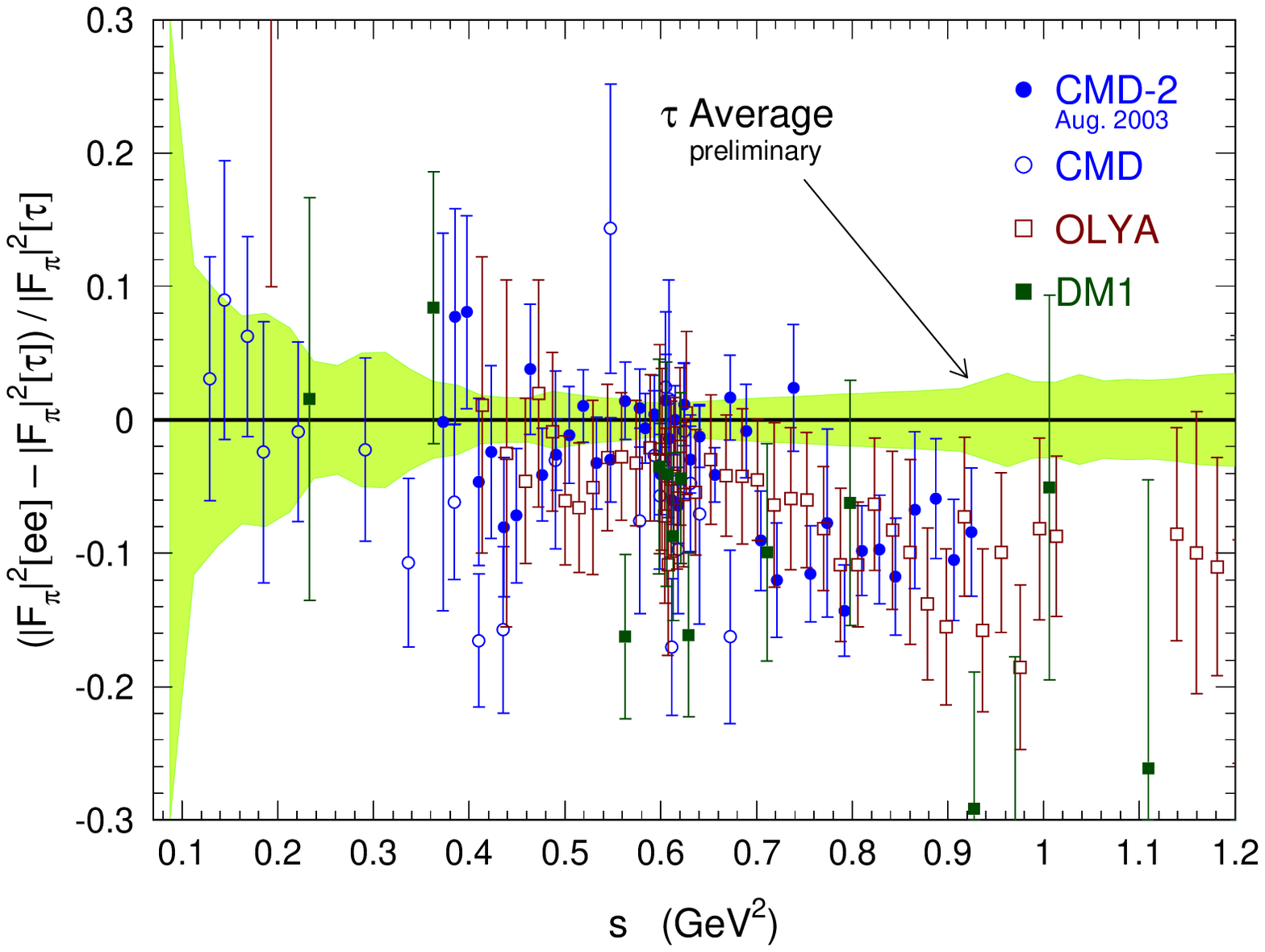, height=5.5cm}
\end{minipage}
\begin{minipage}{0.48\textwidth}
\epsfig{file=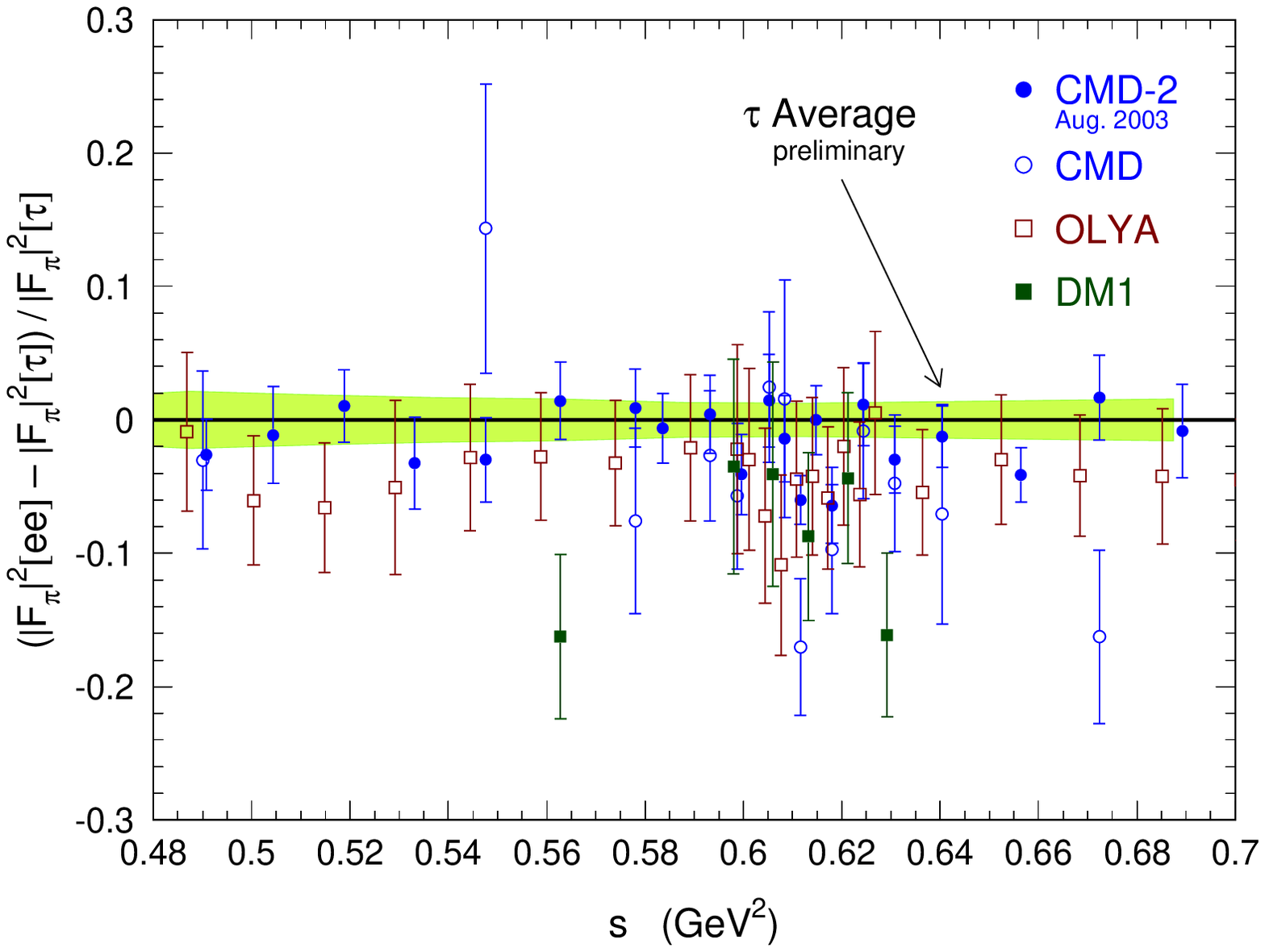, height=5.5cm}
\end{minipage}
\caption{\it The discrepancies between $\tau$ and $e^+e^-$ data at different
energy scales. The data points are $e^+e^-$ data from different experiments,
and the green bands are the $\tau$ data.}
\label{fig:discrepancy}
\end{figure}

The hadronic contribution cannot be calculated from pertubative QCD alone 
because it involves low energy scales near the muon mass. However, the
leading order of hadronic contribution
$a_\mu({\rm had,lo})$\cite{diff1,diff2,diff3}
can be determined directly from the annihilation of $e^+e^-$ to hadrons
through a dispersion integal
\begin{equation}
a_{\mu}({\rm had~lo})=\left( \frac{\alpha m_{\mu}}{3\pi} \right)^2
\int_{4m_{\pi}^2}^{\infty}\frac{ds}{s^2}K(s)R(s),
\label{eq:dispersion}
\end{equation}
in which $s$ is the energy in the center of mass frame, $K(s)$ is a 
kinematic factor and
\begin{equation}
R(s)=\frac{\sigma_{\rm total}(e^+e^-\rightarrow {\rm hadrons})}
	 {\sigma_{\rm total}(e^+e^-\rightarrow \mu^+\mu^-)}.
\end{equation}
The value $R(s)$ can be indirectly determined using data from hadronic
$\tau$ decays, the conserved vector current hypothesis, plus the
appropriate isospin correction. In principle, the large statistics of
$\tau$ decay data could improve the precision of $a_\mu$(had). However,
discrepancies between the $\tau$ and $e^+e^-$ data exist
as shown in Fig.~\ref{fig:discrepancy}.
The $e^+e^-$ data were dominated by the data from CMD2 experiment.
The recent result from radiative return data\cite{kloe} with comparable
uncertainty agree with the CMD2 experiment. The next leading order hadronic
contribution\cite{had21,had22,diff3} and the hadronic light-by-light
scattering\cite{lbl1,lbl2,lbl3,lbl4,lbl5} are calculated by serveral groups
and the combined the results are
\begin{equation}
a_\mu({\rm had~nlo})=10.0(0.6) \times 10^{-10}
\hspace{2cm}
a_\mu({\rm lbl})=12.0(3.5) \times 10^{-10}
\end{equation}

The large discrepancies between $\tau$ and $e^+e^-$ data prevent us from 
combining two to obtain a single standard model prediction. 
The differences between experimental measurement and the standard model
prediction are
\begin{equation}
\Delta a_\mu({\rm e^+e^-})=23.9(7.2_{\rm had~lo})(3.5_{\rm lbl})(6_{\rm exp}) \times 10^{-10}~(2.4\sigma)
\end{equation}
\begin{equation}
\Delta a_\mu(\tau)=7.6(5.8_{\rm had~lo})(3.5_{\rm lbl})(6_{\rm exp}) \times 10^{-10}~(0.9\sigma)
\end{equation}
The question concern the discrepancies betweent $e^+e^-$ and $\tau$ data are
under further theoretical scrutiny and we expected more radiative return
measurements and efforts from latice calculation.

\end{document}